# A THEORY OF QUANTUM GRAVITY MAY NOT BE POSSIBLE BECAUSE QUANTUM MECHANICS VIOLATES THE EQUIVALENCE PRINCIPLE

Mario Rabinowitz
Armor Research
715 Lakemead Way, Redwood City, CA 94062-3922 USA
email: Mario715@gmail.com

## Abstract
Quantum mechanics (QM) clearly violates the weak equivalence principle (WEP).  This implies that quantum mechanics also violates the strong equivalence principle (SEP), as shown in this paper.  Therefore a consistent theory of quantum gravity (QG) that unifies general relativity and QM may not be possible unless it is not based upon the prevailing equivalence principle, or if quantum mechanics can change its mass dependence.  Neither of these possibilities seems likely at the present time.  Examination of QM in n-space, as well as relativistic QM does not change this conclusion.

## 1.  Introduction

Following the 1930 pioneering work of Rosenfeld [1], all attempts to combine Einstein's general relativity (EGR) and quantum mechanics into a theory of quantum gravity (QG) over the past seven decades have been futile -- leading to discrepancies and even contradictions.  A theory of QG may have far-reaching astrophysical implications. Quantum gravity could shed light on the big bang and the early universe. QG may also determine fundamental physical attributes that start on a small scale and affect large scale astrophysical properties.  So it is important to explore a potential incompatibility.  First let us examine the weak and the strong forms of the equivalence principle.   It will be shown that the weak equivalence principle is clearly violated by quantum mechanics.  The violation of the strong equivalence principle by quantum mechanics is more subtle, being difficult to show directly.  However, it will be clearly shown that a violation of the weak equivalence principle implies a violation of the strong equivalence principle upon which EGR is based.

## 2. Quantum Mechanics Clearly Violates the Weak Equivalence Principle

**The weak equivalence principle (WEP) states that there are no physical effects that depend on the mass of a point particle in an external gravitational field.** This is related to the equivalence of inertial and gravitational mass which permits cancellation of the particle's mass in Newtonian physics. Thus the same trajectory is followed by particles of different mass if they have the same initial conditions.

However this does not occur in quantum mechanics. For a particle of inertial mass $m_i$ and gravitational mass $m_g$, falling directly toward mass M where M >> m, the Schroedinger equation is

$$\frac{-\hbar^2}{2m_i}\nabla^2\Psi - \frac{Gm_g M}{r}\Psi = i\hbar \frac{\partial \Psi}{\partial t}. \tag{1}$$

Even for $m_i = m_g$, the mass does not cancel out of the equations of motion, as can be seen more transparently from eq. (2) since m appears in the denominator of the kinetic energy term and the numerator of the potential energy term of the Hamiltonian. This is more apparent for $m_i = m_g \equiv m$, in a uniform gravitational field in the x direction, of acceleration g:

$$\frac{-\hbar^2}{2m}\frac{\partial^2}{\partial x^2}\Psi + mgx\Psi = i\hbar \frac{\partial \Psi}{\partial t}. \tag{2}$$

In examining the possibility of gravitationally bound atoms in 3-space [2] and later in n-space [3, 4] it was clear to me that m remains in the quantized equations of motion, even for M >> m; though m cancels out of the classical equations of motion in Newtonian gravity. One would expect m to cancel out when averaging over states with large quantum numbers that puts them effectively in the classical continuum.

In quantum mechanics, the wavelength is inversely proportional to the momentum and hence involves the mass. Quantum mechanical interference effects in general, and quantum-gravitational interference effects in particular, depend on the phase and phase shifts which depend on the momentum and hence are proportional to the mass. For example for low energy, the phase shift $d_o \propto (k) = (p/\hbar) = mv/\hbar$. This is intrinsic to quantum mechanics. The uncertainty principle is basically related to the destruction of phase relations in different parts of the wave function. The weak equivalence principle requires that the equations of motion be independent of m. Even though the WEP and the SEP work well independently of QM, this is not the case in the union of general relativity and quantum mechanics (quantum gravity).

**3. Quantum Mechanics Violates the Strong Equivalence Principle**

**Einstein postulated the strong equivalence principle (SEP) to formulate general relativity. The SEP states that locally, gravitation is indistinguishable from an equivalently accelerating reference frame.** This implies the WEP since the accelerating reference frame is independent of the mass acted upon by the gravitational field. Furthermore, the SEP implies that it is possible to locally transform away gravitational effects in a properly chosen reference frame.

In the spirit of the SEP by transformation to an accelerated reference frame of acceleration -g, with coordinates $x_a$ in the Schroedinger equation (2), the transformed coordinates

$$x_a = x - vt - \tfrac{1}{2}gt^2 \quad \text{and} \quad t = t_a \tag{3}$$

do not involve the mass m. So m doesn't cancel out, and now both the WEP and the SEP are violated in this case. The violation of the WEP may be less manifest in the Heisenberg matrix quantum mechanics approach, say for example if the QM operator for a particle in a gravitational field were independent of mass. Nevertheless the equations of motion would be mass dependent as the Heisenberg and Schroedinger approaches have been shown to be equivalent. We will soon see that quantum mechanics violates the WEP and the SEP quite generally.

In examining the possible violation of the SEP by quantum mechanics [4], varying modifications of Einstein's SEP were presented. One of these is that of Rohrlich. Interestingly, Rohrlich [5] doesn't use the terms WEP and SEP, and just refers to the equivalence principle (EP). Rohrlich's general statement of EP says that the equations of motion of a nonrotating test body in free fall in a gravitational field are independent of the energy content of that body. He appears to diminish the role of SEP in saying (p.42): "True gravitational fields can never be transformed away. ... Apparent gravitational fields are a characteristic of the motion of the observer (rather than of the observed physical system), while true gravitational fields are the same for all observers no matter what their motion. ... Einstein's statement [of the EP] as an equivalence between accelerated observers and gravitational fields is now seen to be restricted to apparent gravitational fields. True gravitational fields cannot be simulated by acceleration (i.e. by a coordinate transformation)." In my mind, this raises a serious question as to whether "apparent gravitation" and "space-time" are physically synonymous, though they appear to be in EGR. Clearly, the EP does not incorporate spin in general and particle spin in particular. Thus it should not be surprising if the Sagnac [6] effect cannot be properly accounted for within the context of EGR [7].

The different statements of the equivalence principle are interrelated. If the equations of motion involve the mass, m, not only is WEP violated but this also involves a violation of SEP [4].

It is clear that SEP $\Rightarrow$ WEP since SEP yields equations of motion of a body independent of its mass in a gravitational field. A general principle in logic is:

If A $\Rightarrow$ B then (not B) $\Rightarrow$ (not A). (4)

Since SEP $\Rightarrow$ WEP, then (not WEP) $\Rightarrow$ (not SEP). (5)

Therefore quantum mechanics violates the SEP because it violates the WEP.

**4. Semi-Classical Mechanics Violates the Weak Equivalence Principle**

It may not be obvious that the semi-classical Bohr-Sommerfeld condition $\oint p \cdot dl = \oint mv \cdot dl = jh$ violates the WEP. Nevertheless, its predictions for the equations of motion for a particle in a gravitational field clearly do so. We will see that they do so in all dimensions of space.

*4.1 Three Dimensional Space*

Because of the closer proximity and similarity of semi-classical mechanics to classical mechanics, one might think that its violation of the equivalence principle may not be as manifest as that of quantum mechanics. As we shall see, this is not the case.

For a two-body system of masses m << M, the velocity of the orbiting body m is

$$\frac{GmM}{r^2} = \frac{\mathbf{m}v^2}{r} \approx \frac{mv^2}{r} \Rightarrow v = \left(\frac{GM}{r}\right)^{1/2}, \quad (6)$$

where circular motion is considered for simplicity, and the reduced mass $\mathbf{m} = \frac{mM}{m+M} \approx m$ for M>>m. The Bohr-Sommerfeld condition for the angular momentum leads to

$$m(v)r = m\left(\frac{GM}{r}\right)^{1/2} r = j\hbar, \text{ where j is an integer.} \quad (7)$$

Equation (7) implies that the orbital radius is

$$r = \frac{j^2 \hbar^2}{GMm^2}. \quad (8)$$

Substituting eq. (8) into eq. (6) we obtain the velocity

$$v = GMm / j\hbar. \quad (9)$$

and the acceleration is

$$a \equiv v^2 / r = -G^3 M^3 m^4 / (j\hbar)^4. \quad (10)$$

These all have a mass dependence, m, which is not present in their classical counterparts.

### *4.2 n- Dimensional Space*

Sometimes one encounters surprises in higher n-dimensional space [3, 4]: Angular momentum cannot be quantized in the usual manner in four dimensional space. This is because the dependence of angular momentum, L, on $r_n$ allows the orbital radius to adjust in the quantization of L in all dimensions except in 4-space. There is no binding energy for atoms for = 4-space because the binding energy = 0 for n = 4, and the energy levels are all > 0 for n > 4. The macroscopic dimensionality of space can be determined by measuring the temperature dependence of black body radiation because the temperature exponent = n + 1. The volume of a radius r, infinite dimensional sphere = 0 because the n-volume of an n-sphere relative to an n-cube of side = r, peaks ≈ 5-dimensional space. Thereafter for large n, the ratio of the n-sphere volume to the n-cube volume is a quickly diminishing fraction which $\to 0$ as $n \to \infty$. So it is important to look at higher dimensions in case one might encounter a surprise with respect to the WEP.

However, it is easy to demonstrate that semi-classical mechanics violates the WEP in n-dimensional space. Because of the correspondence of the m dependence in quantum mechanics, QM violates the WEP in n-space. We can directly obtain the higher dimensional results for circular orbits directly from [3, 4]:

For M >> m, equating the gravitational force and the centripetal force

$$F_n = \frac{-2\pi G_n M m \Gamma\left(\frac{n}{2}\right)}{\pi^{n/2} r^{n-1}} = \frac{-m v_n^2}{r_n}, \tag{11}$$

where the n-space universal gravitational constant $G_n$ changes, in a way that is model dependent, from its 3-space value. The Gamma function $\Gamma(n) \equiv \int_0^\infty t^{n-1} e^{-t} dt$ for all n (integer and non-integer). When n is an integer, $\Gamma(n) = (n-1)!$ $\hbar$ is (Planck's constant)/$2\pi$. Combining eq. (9) and the Bohr-Sommerfeld condition, $m v_n r_n = j\hbar$, we find for the orbital radius of m around M

$$r_n = \left[\frac{j\hbar \pi^{\frac{n-2}{4}}}{m\left[2G_n M \Gamma(n/2)\right]^{1/2}}\right]^{\frac{2}{4-n}}. \tag{12}$$

In 3-space eq. (12) yields $r_3 = j^2 \hbar^2 / GMm^2$, the same as in eq. (8).

Similarly, the orbital velocity is

$$v_n = \left\{\left[\frac{2\pi G_n M \Gamma\left(\frac{n}{2}\right)}{\pi^{n/2}}\right] \left[\frac{m^{2/(4-n)} \left[2G_n M \Gamma\left(\frac{n}{2}\right)\right]^{\frac{1}{4-n}}}{(j\hbar)^{2/(4-n)} \pi^{(n-2)/2(4-n)}}\right]\right\}^{1/2}. \tag{13}$$

In 3-dimensions eq. (13) gives $v_3 = GMm/j\hbar$, the same result as eq. (9).

The acceleration of the orbiting mass m is

$$a_n = \frac{F_n}{m} = \frac{-2pG_n M \Gamma\left(\frac{n}{2}\right)}{p^{n/2}} \left[\frac{m[2G_n M \Gamma(n/2)]^{1/2}}{j\hbar p^{(n-2)/4}}\right]^{\frac{2n-2}{4-n}}. \tag{14}$$

In 3-dimensional space, eq. (14) yields $a_3 = -G^3 M^3 m^4/(j\hbar)^4$, the same as eq. (10).

It is clear that the acceleration, the orbital radius, and orbital velocity are all functions of the mass m in all dimensions as a result of quantization. The presence of m is not an artifact of the Bohr-Sommerfeld condition. The same mass dependency and basically the same results are obtained from the Schroedinger equation. (The Hamiltonian for the attractive gravitational potential is of the same form as that of the hydrogen atom with radial wave function solutions in terms of Laguerre polynomials. In a one dimensional uniform gravitational potential, the Schroedinger equation can be solved in terms of Airy functions, Bessel functions of order 1/3, or equivalently MacDonald functions.) The failure of m to vanish indicates that quantum mechanics is inconsistent with the weak equivalence principle [2, 3].

The above results indicate that quantum mechanics also violates the WEP in higher dimensional n-space. By the same principle of logic used in Sec. 3, one may conclude that the violation of WEP in n-space implies the violation of SEP in n-space. Thus quantum mechanics violates the SEP in all space and is incompatible with Einstein's general relativity in all space. It is noteworthy that these semi-classical results for gravitational orbits do not approach the classical results as $j \to \infty$ or as $\hbar \to 0$, whereas the quantum mechanical results may have the ability to do so because of additional degrees of freedom in combining wave functions.

**5. Klein-Gordon and Dirac Relativistic Quantum Mechanics Violate the WEP**

The special relativity Klein-Gordon and Dirac equations reduce to the Schroedinger equation in the non-relativistic limit, so their predictions must agree with it in this limit. Thus in the non-relativistic limit, they must give equations of motion that are a function of the particle mass m in a gravitational field, and hence violate the WEP.

The SEP for a non-uniform gravitational field applies at a point, or at best in a very small localized region of space. Therefore in general, the SEP is precluded from applying to QM which is inherently non-local. Since the WEP does not require locality, let us examine these relativistic quantum equations for a free particle to ascertain if they can escape violation of the WEP when a gravitational field is included in the Hamiltonian. Furthermore, a question may be raised about a potential energy representation of gravity since gravity is just space-time curvature in EGR. We avoid the unnecessary complications of a proper depiction of the gravitational source (which in the case of EGR is an energy-momentum tensor that is typically taken to be a kinematic perfect fluid) by considering the free-particle equations.

The Schroedinger equation is inherently not relativistic because it doesn't treat space and time on an equal footing. Its space derivatives are second order and its time derivative is first order. Dirac circumvented this dilemma by modifying the Hamiltonian so that it would be linear in the space derivatives. Thus the Dirac equation for a free mass, m, is

$$\left[ i\hbar \frac{\partial}{\partial t} - i\hbar c \boldsymbol{a} \cdot \nabla + \boldsymbol{b} m c^2 \right] \Psi = 0, \text{ where} \qquad (15)$$

the components of the vector $\boldsymbol{a}$, and the scalar $\boldsymbol{b}$ are Dirac matrices. $\Psi$ is a four-component spinor field. (Incidently, although the Dirac equation is only for special relativity, spinors were discovered by Cartan while working on general relativity.) The matrices $\boldsymbol{a}$ and $\boldsymbol{b}$ are independent of m, as are the first two terms, and $\boldsymbol{b} m c^2 \Psi$ is the only mass term. If one could introduce a special relativistic gravitational potential energy term in the Hamiltonian, it would be proportional to m, and m would not cancel out. Therefore the Dirac equation violates the WEP, implying that it also violates the SEP by the logic of eq. (5), as well as due to non-locality.

The Klein-Gordon equation for a free mass, m, is

$$\left[ \hbar^2 \frac{\partial^2}{\partial t^2} - \hbar^2 c^2 \nabla^2 + m^2 c^4 \right] \Psi = 0. \qquad (16)$$

In addition to a second-order time derivative and a second-order space derivative, the Klein-Gordon equation has a term $m^2c^4\Psi$ which is the only mass term. If one could introduce a special relativistic gravitational potential energy term in the Hamiltonian, it would be proportional to m, and m would not cancel out. Therefore the Klein-Gordon equation violates the WEP, implying that it also violates the SEP by the logic of eq. (5), as well as due to non-locality.

## 6. Discussion

To my knowledge, my approach to the problem preventing the development of a theory of quantum gravity is original and differs from that of others. The only other paper that I was able to find which comes to the same conclusion as mine is that of Loinger [8]. However, he takes a different approach in reaching his conclusion. One difficulty that has been previously examined is the different role that time plays in Einstein's general relativity compared with quantum mechanics. Time is an external scalar parameter in quantum mechanics. In EGR time is part of space-time, and hence is an internal dynamical quantity. It is difficult to reconcile the two concepts. Unruh [9] aptly discusses this in his treatise on *Time, Gravity, and Quantum Mechanics*. Herdegen and Wawrzycki [10] in their paper, *Is Einstein's equivalence principle valid for a quantum particle?* conclude that the EP and quantum mechanics are compatible. Davies [11] discusses tunneling anomalies related to *Quantum mechanics and the equivalence principle.* Although he finds that QM violates the WEP, he does not seem to find an incompatibility between quantum mechanics and the SEP.

## 7. Conclusion

Quantum mechanics clearly violates the WEP, and logic shows that a violation of the WEP implies a violation of the SEP. This is the case for n-dimensional space, as well as for non-relativistic and relativistic QM. Therefore to achieve a consistent theory of quantum gravity, it must not be based upon the prevailing equivalence principle. Or quantum mechanics must change a body's mass dependence (e.g. involving the test mass at high energy and having it cancel out at low energy) in the equations of motion of that body in a gravitational field. Neither of these seems likely at the present time